\documentclass[aps,twocolumn,prd,showpacs,amsmath,amssymb]{revtex4-1}
\usepackage{dcolumn}
\usepackage{bm}
\usepackage{amsfonts}

\newcommand{\ep}{\epsilon}
\newcommand{\be}{\begin{equation}}
\newcommand{\ee}{\end{equation}} 
\newcommand{\eei}{\end{equation}\indent\indent}
\newcommand{\bc}{\begin{center}}
\newcommand{\ec}{\end{center}}
\newcommand{\ber}{\begin{eqnarray*}}
\newcommand{\ear}{\end{eqnarray*}}
\newcommand{\ba}{\begin{array}}
\newcommand{\ea}{\end{array}}

\newcommand{\bea}{\begin{eqnarray}}
\newcommand{\eea}{\end{eqnarray}}

\newcommand{\ei}{\end{itemize}}

\newcommand{\nab}{\nabla}
\newcommand{\la}{\langle}
\newcommand{\ra}{\rangle}

\newcommand \om {\omega}

\def\case#1/#2{\textstyle\frac{#1}{#2} }

\newcommand{\C}{\ensuremath{c_s^2}}
\newcommand{\Ca}{\ensuremath{\frac{dp}{d\mu}}}
\begin{document}

\title{On Shear-Free perturbations of FLRW Universes}

\author{Anne Marie Nzioki$^{ \dag}$, Rituparno Goswami$^{\dag}$,
Peter K.S. Dunsby$^{ \dag \ddag}$ and George F. R. Ellis$^{\dag}$}
\affiliation{\dag \ ACGC and Department of Mathematics and Applied
Mathematics, University of Cape Town, Rondebosch, 7701, South
Africa}
\affiliation{\ddag \  South African Astronomical
Observatory, Observatory, Cape Town, South Africa}

\date{\today}

\email{anne.nzioki@gmail.com, Rituparno.Goswami@uct.ac.za,
George.Ellis@uct.ac.za, Peter.Dunsby@uct.ac.za.}


\begin{abstract}
A surprising exact result for the Einstein Field Equations is that
if pressure-free matter is moving in a shear-free way, then it must
be either expansion-free or rotation-free. It has been suggested
this result is also true for any barotropic perfect fluid, but a
proof has remained elusive. We consider the case of barotropic
perfect fluid solutions linearized about a Robertson-Walker
geometry, and prove that the result remains true except for the case
of a specific highly non-linear equation of state. We argue that
this equation of state is non-physical, and hence the result is true
in the linearized case for all physically realistic barotropic
perfect fluids. This result, which is not true in Newtonian
cosmology, demonstrates that the linearized solutions, believed to
result in standard local Newtonian theory, do not always give the
usual behaviour of Newtonian solutions.
\end{abstract}

\maketitle

\section{Introduction}
This paper deals with a number of interesting properties of
shear-free perfect fluid solutions of General Relativity (GR). The
motivation for this work stems from the desire to probe the
relationship between relativistic and Newtonian cosmology and their
implications for the study of the growth of large-scale structure in
the Universe. Of particular importance is understanding the
differential properties of time-like geodesics which describe the
fluid flow in cosmology. The kinematics of such fluid flows are
described by the expansion $\Theta$, shear (or distortion)
$\sigma_{ab}$, rotation $\om^c$, and acceleration $A_a$ of the
four-velocity field $u^a$ tangent to the fluid flow lines. Their
governing equations are obtained by contracting the Ricci identities
(applied to $u^a$) along and orthogonal to $u^a$, which determine
how they couple to gravity via the Einstein Field Equations
\cite{EllisCovariant}.

Of particular interest is what role the shear plays in the
relationship between Newtonian and relativistic cosmologies. For
example it has been known for some time that quasi-Newtonian
descriptions of cosmology, the so-called {\em Silent models}, may be
constructed for observers which move along geodesics which are both
shear-free and irrotational \cite{silent}. The intricate
relationship between the kinematic quantities in Newtonian and
relativistic cosmologies is most strikingly seen in a remarkable
result first obtained by one of us in 1967 \cite{Ellis}. In this
paper it was found that {\em if the four velocity vector field of a
barotropic perfect fluid with vanishing pressure is shear free, then
either the expansion or the rotation of the fluid vanishes}. This is
a purely local result to which no corresponding Newtonian equivalent
appears to hold, as counter-examples can be explicitly constructed
\cite{counter}. Given that this theorem and its extensions appear to
hold for arbitrarily weak fields and for fluids of arbitrarily low
density, one needs to understand why the Newtonian approximation
fails.

The result has be extended to general barotropic fluids for number
of special cases by Senovilla \cite{SSS}, but has yet to be proved
in general. As a first step towards this goal, we examine what ever
result holds in situations where the hydrodynamic and gravitational
equations have been linearised. Of course there are many ways of
doing this, but one way that is cosmologically relevant is to
linearise the equations about a
Friedmann-Lema\^{\i}tre-Robertson-Walker (FLRW) background
\cite{EB,EBH,BDE,DBE,BED}. These {\em almost FLRW} models can be
thought of as lying somewhere between the full non-linear GR
situation and Newtonian theory, at least in the cosmological
context, and therefore an analysis of theorem in this context could
shed some light on the generality of the result. We show that it
remains true for such linearised barotropic perfect fluid solutions,
unless the fluid obeys a highly non-linear equation of state (see
(\ref{DE}) below) which we argue is non-physical. Hence the result
remains true for physically realistic equations of state in an
almost-FLRW geometry.

This result will be useful in obtaining and studying new
perfect-fluid solutions of Einstein's field equations with a
shear-free velocity vector field, and in examining how linearized
General Relativity solutions relate to the Newtonian case, which is
the foundation of astrophysical studies in cosmology.

\section{Linearised Field equations about FLRW background}

To perturb the FLRW spacetime we use the standard 1+3 covariant
approach \cite{EllisCovariant}, where we must first define a
time-like congruence with a unit tangent vector $u^a$. The natural
choice of this vector 
is tangent to the the matter flow lines. Then the spacetime is split
locally in the form $R\otimes V$ where  $R$ denotes the worldline
along $u^a$ and  $V$ is the 3-space perpendicular to $u^a$. Then any
vector $X^a$  can be projected on the 3-space by the projection
tensor $h^a{}_b=g^a{}_b+u^au_b$.  At this point, two derivatives are
defined: the vector $ u^{a}$ is used to define the \textit{covariant
time derivative} along the observers' worldlines(denoted by a dot)
for any tensor $ T^{a..b}{}_{c..d}$, given by \be
\dot{T}^{a..b}{}_{c..d}{} = u^{e} \nab_{e} {T}^{a..b}{}_{c..d} \ee
and the tensor $ h_{ab} $ is used to define the fully orthogonally
\textit{projected covariant derivative} $D$ for any tensor $
T^{a..b}{}_{c..d} $: \be D_{e}T^{a..b}{}_{c..d}{} = h^a{}_f
h^p{}_c...h^b{}_g h^q{}_d h^r{}_e \nab_{r} {T}^{f..g}{}_{p..q}\;,
\ee with total projection on all the free indices.  Angle brackets
denote orthogonal projections of vectors, and the orthogonally
\textit{projected symmetric trace-free} PSTF part of tensors: \be
V^{\la a \ra} = h^{a}{}_{b}V^{b}~, ~ T^{\la ab \ra} = \left[
h^{(a}{}_c {} h^{b)}{}_d - \frac{1}{3} h^{ab}h_{cd}\right] T^{cd}\;.
\label{PSTF} \ee This splitting of spacetime also naturally defines
the 3-volume element \be \ep_{a b c}=-\sqrt{|g|}\delta^0_{\left[ a
\right. }\delta^1_b\delta^2_c\delta^3_{\left. d \right] }u^d\;,
\label{eps1} \ee with the following identities \be \ep_{a b c}\ep^{d
e f}=3!h^d_{\left[ a \right. }h^e_bh^f_{\left. c \right] }\;\;;\;\;
\ep_{a b c}\ep^{d e c}=2!h^d_{\left[ a \right. }h^e_{\left. b
\right] } \label{eps2}. \ee

The covariant derivative of the time-like vector $u^a$ can now be
decomposed into the irreducible parts as \be
\nabla_au_b=-A_au_b+\frac13h_{ab}\Theta+\sigma_{ab}+\ep_{a b
c}\om^c\;, \ee where $A_a=\dot{u_a}$ is the acceleration,
$\Theta=D_au^a$ is the expansion, $\sigma_{ab}=D_{\la a}u_{b \ra}$
is the shear tensor and $w^a=\ep^{a b c}D_bu_c$ is the vorticity
vector. Similarly the Weyl curvature tensor can be decomposed
irreducibly into the Gravito-Electric and Gravito-Magnetic parts as
\be E_{ab}=C_{abcd}u^cu^d=E_{\la ab\ra}\;;\;
H_{ab}=\frac12\ep_{acd}C^{cd}_{be}u^e=H_{\la ab\ra}\;, \ee which
allows a covariant description of tidal forces and gravitational
radiation.

In the 1+3 covariant perturbation theory
\cite{EB,EBH,BDE,DBE,BED,Roy,DBBE,Roy2}, we consider  the background
to be FLRW where the Hubble scale sets the scale for the
perturbations. The quantities that vanish in the background
spacetime are considered to be first order and are automatically
gauge-invariant by virtue of the Stewart and Walker lemma \cite{SW}.
In the perturbed spacetime the matter is considered to be a perfect
fluid with the Energy Momentum tensor \be
T^{ab}=(\mu+p)u^au^b+pg^{ab}\;, \label{EMT} \ee so the vector field
$u^a$ is uniquely defined as the timelike eigenvector of $T^{ab}$ as
long as $\mu+p\neq 0$ (the heat flux $q^a$  and the anisotropic
stress $\pi_{a b}$ vanish in the perturbed spacetime). Furthermore,
we assume the matter to have a barotropic equation of state
$p=p(\mu)$ satisfying the Weak and Dominant energy conditions. We
exclude the vacuum case, therefore the energy conditions will be \be
\mu>0\;\;;\;\;\mu+p>0\;\;;\;\;\mu\ge|p|\; \label{EC} \ee for both
the background spacetime and the perturbed solution (and the
Minkowski and De Sitter backgrounds will not occur). The local
isentropic sound speed is \be c_s^2\equiv \frac{dp}{d\mu}\;\;;\;\;
0\le\C\le1\;. \label{SS}\ee The bound on the local sound speed is
required for local stability of matter (lower bound) and causality
(upper bound), respectively.

Now we consider shear-free perturbations and hence the shear tensor
($\sigma_{a b}$) vanishes identically. With the conditions above,
the linearised  field equations are then as follows:
\subsection*{Propagation equations}
\be
\dot{\Theta}=D_aA^a-\frac13 \Theta^2-\frac12(\mu+3p)\;,
\label{R1}
\ee
\be
\dot{\om}^{\la a \ra}=\frac12\ep^{abc}D_bA_c-\frac23\Theta\om^a\;,
\label{R3}
\ee
\be
\dot{H}^{\la ab \ra}=-\ep^{cd\la a}D_cE^{\ra b}_d-\Theta H^{ab}\;,
\label{B2}
\ee
\be
\dot{E}^{\la ab \ra}=\ep^{cd\la a}D_cH^{\ra b}_d-\Theta E^{ab}\;,
\label{B1}
\ee
\be
\dot{\mu}=-\Theta(\mu+p)\;,
\label{B4}
\ee
\subsection*{Constraint equations}
\be (C_0)^{ a b}:=E^{a b}-D^{\la a}A^{b \ra}=0\;, \label{R2} \ee \be
(C_1)^a:=D^a\Theta-\frac32\ep^{abc}D_b\om_c=0\;, \label{R4} \ee \be
(C_2):=D^a\om_a=0\;, \label{R5} \ee \be (C_3)^{ a b}:=H^{a b}+D^{\la
a}\om^{b \ra}=0\;. \label{R6} \ee \be (C_4)^a:=D^ap +
(\mu+p)A^a=0\;, \label{B3} \ee \be (C_5)^a:=D_bE^{a
b}-\frac13D^a\mu=0\;, \label{B5} \ee \be (C_6)^a:=D_bH^{a
b}+(\mu+p)\om^a=0\;. \label{B6} \ee We note that the constraints
$(C_1)^a$, $(C_2)$, $(C_3)^{a b}$, $(C_5)^a$ and $(C_6)^a$  are the
constraints of the Einstein field equations for general matter
motion specialized to the shear-free case and are known to be consistently {\it time propagated} along
$u^a$ locally. However the conditions  $\sigma_{a b}=0$  and $q^a=0$
give the two new constraints $(C_0)^{a b}$ and $(C_4)^a$
respectively.

We also use the following linearised commutation relations for shear-free congruences:
For any scalar `$f$'
\bea
[D_aD_b-D_bD_a]f&=&2\ep_{a b c}\om^c\dot f \;, \nonumber\\
\ep^{a b c}D_bD_c f&=&2\om^a \dot f\;.
\label{C1}
\eea
If the gradient of the scalar is of the first order, we then have
\bea
[D^aD_bD_a-D_bD^2]f&=&\frac23\left(\mu-\frac13\Theta^2\right)D_bf\;,
\label{C2}
\eea
\bea
[D^2D_b-D_bD^2]f&=&\frac23\left(\mu-\frac13\Theta^2\right)D_bf\nonumber\\
&&+2\ep_{dbc}D^d(\om^c\dot f)\;. \label{C3} \eea Also for any first
order 3-vector $V^a=V^{\la a \ra}$, we have \bea
[D^aD_b-D_bD^a]V_a&=&\frac23\left(\mu-\frac13\Theta^2\right)
h^a{}_{\left[a\right. } V_{\left. b \right]}\;, \label{C4} \eea \bea
h^a{}_ch^d{}_b(D_dV^c)\dot{}=D_b\dot{V^{\la a \ra}}-\frac13\Theta
D_bV^a \label{C6} \eea \bea h^a{}_c(D^2V^c)\dot{}=D_b(D^{\la b}V^{a
\ra})\dot{}-\frac13\Theta D^2V^a\;. \label{C5} \eea Using the field
equations and identities of this section we will now investigate the
compatibility of the new constraints with the existing ones in terms
of the consistency up to the linear order of their spatial and
temporal propagation.

\section{Consistency of the new constraints}

The conditions of shear-free perturbations and the matter being a
perfect fluid in the perturbed spacetime give the new constraints
$(C_0)^{a b}$ and $(C_4)^a$ respectively. To check their
compatibility with the linearised existing constraints of Einstein
field equations (henceforth all the equations are up to the linear
order), we plug $(C_0)_{b d}$ in $(C_5)_b$ to get \be D^dD_{\la
b}A_{d \ra}-\frac13D_b\mu=0\;. \label{subs1} \ee Now from the
constraint $(C_4)_b$ we have \be A_b=-\frac{c_s^2}{\mu+p}D_b\mu
\label{subs2} \ee Using equation (\ref{subs2}) in (\ref{subs1}) we
get the constraint \be (C_7)_b:= \frac{\C}{\mu+p}D^dD_{\la b}D_{d
\ra}\mu+\frac13D_b\mu=0\;. \label{newcons} \ee 
For the new constraints $(C_0)^{a b}$ and $(C_4)^a$ to be compatible
with the existing ones, the constraint $(C_7)_b$ must be satisfied.

To check the spatial consistency of $(C_7)_b$ on any initial
hypersurface we take the curl of (\ref{newcons}) to get \be
\frac{\C}{\mu+p}\ep^{acb}D_c D^dD_{\la b}D_{d
\ra}\mu+\frac13\ep^{acb}D_cD_b\mu=0\;, \label{curl1} \ee which using
(\ref{C1}) gives \be \frac{\C}{\mu+p}\ep^{acb}D_c D^dD_{\la b}D_{d
\ra}\mu+\frac23\om^a\dot{\mu}=0\;. \label{curl2} \ee Breaking the
PSTF part according to equation (\ref{PSTF}) and using the
commutators (\ref{C2}), (\ref{C3}) we have \bea
\frac{\C}{\mu+p}\ep^{acb}\left[\frac23D_cD_bD^2\mu+\frac23\left(\mu-\frac13\Theta^2\right)
D_cD_b\mu\right. \nonumber\\
+\left. \dot{\mu}\ep_{dbk}D_cD^d\om^k\right]+\frac23\om^a\dot{\mu}=0\;.
\eea
Again using (\ref{C1}) and (\ref{eps2}) in the above equation
we get
\bea
\frac{\C}{\mu+p}\left[\frac43\left(\mu-\frac13\Theta^2\right)\om^a\dot{\mu}-\dot{\mu}D_kD^a\om^k
\right. \nonumber\\
+ \left. \dot{\mu}D^2\om^a\right]+\frac23\om^a\dot{\mu}=0\;.
\label{curl3} \eea Now from the relation (\ref{C3}) and using
(\ref{R5}) we know \be
D_kD^a\om^k=\frac23\left(\mu-\frac13\Theta^2\right)\om^a\;,
\label{curl4} \ee Plugging (\ref{curl4}) and (\ref{B4}) in
(\ref{curl3}) and simplifying we finally get \be (C_8)^a : =
\Theta\left[\frac23\om^aY+\C D^2\om^a\right]=0\;, \label{Result1}
\ee where \be Y=\mu+p+\C\left(\mu-\frac13\Theta^2\right)\;.
\label{Y} \ee

From $(C_8)^a$ we can immediately see that for matter with constant
pressure
 ($p={\rm{constant}}\,\Rightarrow\,\C=0$), shear-free
perturbations are consistent iff $\Theta \om^a=0$ (as according to
the second condition of (\ref{EC}), $\mu+p>0$). That is, if the
geodesics of the matter congruence in  the perturbed spacetime are
shear-free then they should be either expansion-free or
vorticity-free (or both). This shows that {\em the results of
\cite{Ellis} and \cite{SSS} for pressure-free matter are true for
the linearized theory.}

However for a general equation of state, all we can say from the
equation (\ref{Result1}) is, either the matter congruence is
expansion free ($\Theta=0$), or the vorticity vector must satisfy
\be (C_9)^a:= \frac23\om^aY+\C D^2\om^a=0\;, \label{newcons1} \ee
for the new constraints to be spatially consistent on any initial
hypersurface.

Now let us check the temporal consistency of the constraint (\ref{newcons1}).
Propagating it along $u^a$ we get
\be
(\C D^2\om^a)\dot{} + \frac23(\om^aY)\dot{}=0\;.
\label{prop1}
\ee
We can easily see that
\be
\dot{\C}=-\Theta(\mu+p)\frac{d^2p}{d\mu^2}\;.
\label{prop2}
\ee
Now from (\ref{C5})we have
\bea
\C(D^2\om^a)\dot{}=\C[D_b(D^{\la b}\om^{a \ra})\dot{}-\frac13\Theta D^2\om^a]\;.
\label{prop2a}
\eea
We know from the constraint (\ref{R5}) that
\bea
D_b(D^{\la b}\om^{a \ra})\dot{}=\frac12 D_b[(D^b\om^a)\dot{}+(D^a\om^b)\dot{}]\;.
\label{prop2b}
\eea
Using (\ref{C6}) the equation (\ref{prop2b}) becomes
\bea
D_b(D^{\la b}\om^{a \ra})\dot{}=\frac12 D_b\left[D^b\dot{\om^{\la a \ra}}-
\frac13\Theta D^b\om^a\right.
\nonumber\\
\left. +D^a\dot{\om^{\la b \ra}}-\frac13\Theta D^a\om^b\right]\;.
\label{prop2c}
\eea
Simplifying the above equation using (\ref{R3}), (\ref{B3}) and (\ref{C1}), we get
\bea
D_b(D^{\la b}\om^{a \ra})\dot{}=-\frac12\Theta(1-\C)(D^2\om^a+D_bD^a\om^b)\;.
\label{prop2d}
\eea
Putting equation (\ref{prop2d}) in (\ref{prop2a}), we have
\bea
\C(D^2\om^a)\dot{}=-\Theta\alpha\C D^2\om^a-\Theta\beta D_bD^a\om^b\;,
\label{prop3}
\eea
where
\bea
\alpha=-\frac\C2+\frac56\;\;;\;\; \beta=\frac\C2(1-\C)\;.
\eea
Using (\ref{newcons1}) and (\ref{curl4}), (\ref{prop3}) becomes
\bea
\C(D^2\om^a)\dot{}=\frac23\om^a\Theta\left[\alpha Y-\beta\left(\mu-\frac13\Theta^2\right)\right]
\;.
\label{prop4}
\eea
Combining (\ref{prop2}) and (\ref{prop4}) and using (\ref{newcons1}) we get
\bea
(\C D^2\om^a)\dot{}=\frac23\om^a\Theta\left[\frac{Y}{\C}(\mu+p)\frac{d^2p}{d\mu^2}+
\alpha Y\right.\nonumber\\
\left.-\beta\left(\mu-\frac13\Theta^2\right)\right]\;.
\label{prop5}
\eea
Also from  (\ref{R1}), (\ref{B4}) and (\ref{prop2}) we have
\bea
\dot{Y}=-\Theta\left[(\mu+p)\left(\mu-\frac13\Theta^2\right)\frac{d^2p}{d\mu^2}+Z\right]\;,
\label{prop6}
\eea
where
\bea
Z=(\mu+p)(1+\C)+\frac23\C\left(\mu-\frac13\Theta^2\right)\;.
\eea
Now using (\ref{R3}), (\ref{B3}), (\ref{C1}) and (\ref{prop6}) we get
\bea
\frac23(\om^aY)\dot{}=-\frac23\om^a\Theta\left[\left(-\C+\frac23\right) Y \right. \nonumber\\
\left.
+(\mu+p)\left(\mu-\frac13\Theta^2\right)\frac{d^2p}{d\mu^2}+Z\right]
\label{prop7} \eea 
Finally using (\ref{prop5}) and (\ref{prop7}) in
(\ref{prop1}) and simplifying, we get \bea 
\frac23\om^a\Theta(\mu+p)\left[(\mu+p)\frac{d^2p}{d\mu^2}-\C\left(\frac56+\frac\C2\right)
\right.\nonumber\\
\left.-\,\frac{{}^3R}{2(\mu+p)}c_s^4\left(1-\C\right)\right]=0\;.
\label{final} \eea where ${}^3R=2[\mu-(1/3)\Theta^2]$ is the spatial
curvature. In FLRW spacetimes it can be written in term of the scale
factor `$a(t)$' as, \bea
{}^3R=\frac{k}{a(t)^2}=k\exp\left\{\frac23\int\frac{d\mu}{\mu+p}\right\}\;,
\label{3curv} \eea where $k=-1,0,+1$ denotes open, flat and closed
universes respectively. Thus we can easily see that for the new
constraints to be spatially and temporaly consistent we must have
either $\om^a\Theta=0$ or the barotropic equation of state must
satisfy the following non-linear higher order DE: \bea
(\mu+p)\frac{d^2p}{d\mu^2}-\Ca\left(\frac56+\frac12\Ca\right)\,\,\,\,\nonumber\\
-k\frac{
\exp\left\{\frac23\int\frac{d\mu}{\mu+p}\right\}}{2(\mu+p)}\left(\Ca\right)^2\left(1-\Ca\right)=0\;.
\label{DE} \eea

We see that {\em the shear-free results of \cite{Ellis} and
\cite{SSS} are avoided, at least at the linearised level, if the
equation of state of the matter solves (\ref{DE})}. However, {\em a
priori} it seems highly unlikely that any realistic barotropic
equation of state will obey this extremely non-linear equation. We
now try to find solutions of this equation, under various simplified
assumptions or realistic initial conditions, to confirm it is
nonphysical.

\begin{enumerate}
\item \underline{Flat universe ($k=0$) with $\C={\rm constant}\ne0$:}

This is the simplest case in which the equation (\ref{DE}) reduces
to a simple algebraic equation \be
\left(\frac56+\frac12\C\right)=0\;, \ee which gives $\C=-5/3$. This
is physically not possible as the lower bound on the local sound
speed (\ref{SS}) 
is violated.

\item \underline{Closed/open universe with $\C={\rm constant}\ne0$:}
In this case also, the equation (\ref{DE}) reduces to an algebraic
equation, and we get the relation \bea
{}^3R=-2\frac{\left(\frac56+\frac12\C\right)}{\C\left(1-\C\right)}(\mu+p)
\label{DE1} \eea Differentiating (\ref{DE1}) with respect to $\mu$
and using (\ref{3curv}) we get \bea
\frac23\frac{{}^3R}{(\mu+p)}=-2\frac{\left(\frac56+\frac12\C\right)}{\C\left(1-\C\right)}
(1+\C)\;. \label{DE2} \eea Eliminating ${}^3R/(\mu+p)$ from
(\ref{DE1}) and (\ref{DE2}) we get the solution $\C=-1/3$, which
again violates the lower bound of the local sound speed.

\item  \underline{Flat universe with varying sound speed:}
In this case the equation (\ref{DE}) becomes \bea
(\mu+p)\frac{d^2p}{d\mu^2}-\Ca\left(\frac56+\frac12\Ca\right)=0\;.
\label{DE3} \eea To solve (\ref{DE3}), if we choose the initial
epoch ($\mu=\mu_0$) to be a radiation dominated one (which is quite
realistic in view of our current understanding of the universe) with
$\C\approx 1/3$, then from (\ref{DE3}) we can easily see that $\C$
monotonically increases with $\mu$. And in the interval
($\mu_0\le\mu<\infty$) the function $p(\mu)$ is concave upwards.
Therefore there must exist an earlier epoch at which $p(\mu)>\mu$,
which violates (\ref{EC}).

\item\underline{Closed/open universe with varying sound speed:}
This is the most general case and let us try to find a solution with
similar initial conditions as in the previous case. Since we know
that very early universe was radiation dominated, let us suppose
that there exists an epoch ($a_0<<1$) with density $\mu_0$ and
pressure $p_0$ where $(\mu_0,p_0)\approx 1/a_0^4$. As we have
already seen, ${}^3R\approx 1/a_0^2$, hence the last term on the LHS
of (\ref{DE}) becomes suppressed  and in this case also one can
easily show that $\C$ monotonically increases with $\mu$. Therefore
there must exist an earlier epoch $a_1<a_0$ with $\mu_1>\mu_0$,
where $p(\mu)>\mu$ and (\ref{EC}) is violated. In other words, no
solution satisfying (\ref{EC}) exists for (\ref{DE}) that gives a
radiation dominated era in the early universe.
\end{enumerate}

Hence for any physically realistic barotropic equation of state, if the new constraints are to be consistently
propagated, we must have
$\om^a\Theta=0$. We thus proved an important theorem for shear-free perturbations
of FLRW spacetimes:

{\em For an ``almost'' homogeneous and isotropic Universe filled
with a barotropic perfect fluid subject to a physically realistic
equation of state, if the fluid congruence is shear-free in a domain
U, then it must be either vorticity-free or expansion-free in U.}

\section{Discussions}

This result gives an interesting scenario. The linearised shearfree
solutions - almost universally used to study the formation of
structure by gravitational instability in the expanding universe,
and believed to result in standard local Newtonian theory - do not
have the same behaviour as shearfree Newtonian solutions. This may
affect simple structure formation scenarios for rotating matter.

Another interesting point that emerged from our analysis is that
there exists a class of barotropic equation of state (however
unphysical that may be) for which the usual shear-free result can be
avoided in the linearised case. It would be an interesting problem
to see whether this same class of equations of state (or some
similar class) allows shear free rotating and expanding solutions
for the full non-linear Einstein equations for a barotropic perfect
fluid.


\end{document}